\documentclass[aastex]{emulateapj}

\usepackage{graphicx,longtable}
\usepackage{array}
\usepackage{amsmath}
\usepackage{amssymb}
\usepackage{subfigure}
\usepackage{color}

\shorttitle{Relative contributions to the $r$-process}
\shortauthors{Shibagaki et al.}

\begin{document}
\title{Relative contributions   of the weak, main and  fission-recycling $r$-process}

\author{S. Shibagaki$^{1,2}$, {T.Kajino$^{2,1}$, G. J. Mathews$^{3,2}$, S. Chiba$^{4,2}$, S. Nishimura$^{5,2}$, G. Lorusso$^{6,5,7}$}}

\altaffiltext{1}{Department of Astronomy, The University of Tokyo, 113-033 Tokyo, Japan}

\altaffiltext{2}{National Astronomical Observatory of Japan, 2-21-1
Osawa, Mitaka, Tokyo, 181-8588, Japan\\}

\altaffiltext{3}{Center for Astrophysics, Department of Physics, University of Notre Dame, Notre Dame, IN 46556, USA\\}

\altaffiltext{4}{Research Laboratory for Nuclear Reactors, Tokyo Institute of Technology,  2-12-1 Ookayama, Meguro, Tokyo 152-8550, Japan\\}

\altaffiltext{5}{RIKEN Nishina Center,  2-1 Hirosawa, Wako-shi, Saitama 351-0198, Japan\\}

\altaffiltext{6}{National Physical Laboratory, Teddington, Middlesex TW11 0LW, United Kingdom\\}

\altaffiltext{7}{Department of Physics, University of Surrey, Guildford GU2 7XH, United Kingdom\\}

\begin{abstract}
There has been a persistent conundrum in attempts to model the nucleosynthesis of heavy elements by rapid neutron capture (the $r$-process).  Although the location of the abundance peaks near nuclear mass numbers 130 and 195 identify an environment of rapid neutron capture near closed nuclear shells, the abundances of elements just above and below those peaks are often underproduced by more than an order of magnitude in model calculations.  At the same time there is a debate in the literature as to what degree  the $r$-process elements are produced in supernovae or the mergers of binary neutron stars.  In this paper we propose a novel solution to both problems.  We demonstrate that the underproduction of  nuclides above and below the $r$-process peaks in  main or weak $r$-process models (like magnetohydrodynamic jets or neutrino-driven winds in core-collapse supernovae) can be supplemented via fission fragment distributions from the  recycling of material in a neutron-rich environment such as that encountered in neutron star mergers.  In this paradigm, the abundance peaks themselves are well reproduced by a moderately neutron rich, main $r$-process  environment such as that encountered in the magnetohydrodynamical jets in supernovae supplemented with a high-entropy, weakly neutron rich environment such as that encountered in the neutrino-driven-wind model to produce the lighter $r$-process isotopes.    Moreover, we show that the relative contributions to the $r$-process abundances in both the solar-system and metal-poor stars from the weak, main, and fission-recycling environments required by this proposal are consistent with estimates of the relative Galactic event rates of core-collapse supernovae for the weak and main $r$-process and neutron star mergers for the fission-recycling $r$-process.  

\end{abstract}
\medskip

\keywords{nuclear reactions, nucleosynthesis, abundances - supernovae: general - stars: abundances}

\maketitle

\vskip 1.3cm



\section{Introduction}

It has been known for more than half a century \citep{Burbidge57} that about half of the elements heavier than iron are produced via rapid neutron capture (the $r$-process).
Indeed, the basic physical conditions for the $r$-process are well constrained \citep{Burbidge57} by simple nuclear physics.
The observed abundance distribution requires  a sequence of near equilibrium rapid neutron captures
and photoneutron emission reactions far on the neutron-rich side of stability.
This equilibrium is established with a maximum abundance strongly peaked
on one or two isotopes far from stability.
The relative abundance of $r$-process elements is then determined
by the relative $\beta$-decay rates along this $r$-process path.,
i.e., longer $\beta$-decay lifetimes result in higher abundances.

In spite of this simplicity, however, the unambiguous identification of the site for the $r$-process nucleosynthesis has remained elusive.  Parametrically, one can divide current models for the $r$-process into  three scenarios roughly characterized by the number of neutron captures per seed nucleus ($n/s$).  This parameter, in turn is the consequence of a variety of conditions such as time-scale, baryon density, average charge per baryon, $Y_e \equiv \langle Z/A\rangle$, and entropy (or baryon to photon ratio) corresponding to different astrophysical environments \citep[e.g.][]{Meyer97, Otsuki03}.   

An environment in which there are few neutron captures per seed ($n/s \sim 50$) produces what has been identified as the weak $r$-process \citep{Wasserburg96}.  It can only produce the lightest $r$-process nuclei up to $A \sim 125$.  Such an environment may  occur, for example, in the neutrino-driven wind of core-collapse supernovae  (CCSNe) \citep{Woosley94,Wanajo13},   part of the outflow from the remnant of compact binary  mergers \citep{Rosswog14,Perego14,Just15},  the delayed magnetohydrodynamic (MHD) jet from CCSNe \citep{Nishimura15}.

An environment with enough neutron captures per seed ($n/s \sim 100$)  to produce the two $r$-process abundance peaks at $A = 130$ and 195, corresponds to what has been dubbed the main $r$-process and could correspond, for example, to the ejection of neutron-rich material via magnetic  turbulence in magnetohydrodynamically driven jets (MHDJ)  from core collapse supernovae \citep{Nishimura06,Fujimoto07,Fujimoto08,Ono12,Winteler12,Nakamura14,Nishimura15}, or in neutron star mergers (NSMs) \citep{Wanajo14,Goriely15}.  

In this paper, we are particularly  interested in a third environment that  we dub the fission-recycling $r$-process.  In this environment the number of neutron captures per seed nucleus can be very large ($n/s \sim 1000$).  The $r$-process path then proceeds along the neutron drip line all the way to the  region of fissile nuclei ($A \approx 300$) where the $r$-process is terminated by beta- or neutron-induced fission.  Fission recycling can then occur whereby the fission fragments continue to experience neutron captures until  beta- or neutron-induced fission again terminates the $r$-process path.  After a few cycles the abundances can become dominated by the fission fragment distributions and not as much by the beta-decay flow near the closed shells.  Hence, a very different mass distribution can ensue.  Such environments are often associated with {the dynamical ejecta from } NSMs in which the  tidal ejection of neutron-rich material during the merger can lead to fission recycling and many neutron captures per seed \citep[e.g.][]{Goriely11, Korobkin12, Piran13, Rosswog13,Goriely13}.

Observations \citep{Sneden08} showing the appearance of heavy-element $r$-process abundances early in the history of the Galaxy seem to favor the short progenitor  lifetime of CCSNe over NSMs as the $r$-process site.  However, identifying the $r$-process site in models of CCSNe has been difficult \citep{Arnould07, Thielemann11}.   

The three types of environments, neutrino-driven winds (NDWs),
magnetohydrodynamic jets (MHDJs), and neutron star mergers (NSMs), have
all been studied extensively in the recent literature. The robustness of the results
varies for different environments, but uncertainties in both
astrophysical conditions and nuclear input are well recognized in all
cases.  For example, the  previously popular model \citep{Woosley94} of $r$-process nucleosynthesis in the NDW  above the newly forming neutron star has been shown \citep{Fischer10,Hudepohl10} to be inadequate as a main $r$-process site when modern neutrino transport methods have been employed.  
The  required neutron captures per seed do not  occur in the neutrino energized wind. Nevertheless, it is quite likely that the weak $r$-process does occur \citep{Wanajo13} in the NDW producing neutron rich nuclei up to about $A \sim 125$.  

Regarding the weak $r$-process, however, one should note that while most $s$-process models now produce r-process residuals for  $A _\sim^\ge 120$ which look remarkably similar, this is not the case for lighter nuclei.  Hence, there is some uncertainty in determining the solar-system $r$-process abundances via subtraction of the $s$-process contribution from the total abundances.  Indeed, a final consensus has not yet been reached on the predicted $s$-process abundances of light elements.  Hence,  what is usually taken to be  the weak $r$-process [or unknown lighter element primary process (LEPP)] may actually correspond to a much different environment.  For example in  \cite{Trippella14} it was demonstrated that enhanced   light-element abundances  could arise via non-parametric MHD-driven mixing mechanisms.  This enhanced light-element  $s$-process component could obfuscate the need for a  weak $r$-process.  Nevertheless, with this {\it caveat} in mind we adopt the NDW in supernovae as representative of the weak $r$-process.  We note, however, that our arguments below regarding the relative contribution of the weak $r$-process may in fact refer to the relative contribution of an $s$-process driven LEPP.   

Indeed, the difficulties in reproducing the $r$-process abundances have motivated many new studies of NSMs \citep[e.g.][]{Goriely11, Korobkin12, Goriely13, Wanajo14, Perego14}.  Nevertheless, one scenario for the $r$-process in CCSNe remains viable.  It is the MHDJ model \citep{Nishimura06,Fujimoto07,Fujimoto08,Ono12,Winteler12,Nakamura14,Nakamura15,Nishimura15}.  In this model magnetic turbulence leads to the ejection of neutron rich material into a jet.  As the jet transports this  neutron-rich  material away from the star it can undergo $r$-process  nucleosynthesis in a way that avoids the low neutron-to-seed ratios  associated with neutrino interactions in the  NDW model.  Moreover, the required conditions of the $r$-process environment
(timescale, neutron density, temperature, entropy, electron fraction, etc.) are well accommodated in this model.

However, there is a persistent problem in this model, or any general model \citep[e.g.][]{Meyer97, Otsuki03} in which the $r$-process elements are produced in a short time scale via the rapid expansion of material away from a neutron star.  In such models, the neutron density rapidly diminishes and $r$-process path freezes out near the neutron closed shells far from stability.  Most such models underproduce isotopic abundances just below and above the $r$-process abundance peaks as we describe in more detail below.

 Indeed, all $r$-process models are fraught with uncertainties in the input nuclear physics, the astrophysical environment, and the galactic chemical evolution.  Rather than to give up, however, it is highly desirable to explore any possible method in which the relative contributions of each of the primary environments (weak, main, and fission recycling) could be ascertained from observation.  In this paper we propose such a possibility.

With this in mind our goal  is to analyze the general advantages and disadvantages of each of the characteristic  environments.  Although we have noted here specific astrophysical models that are likely to be associated with the various  conditions, the
readers should be aware of the uncertainties involved and consider these environments 
 as illustrative, not definitive.   Nevertheless, we speculate  here that it may be possible to quantify the relative contribution of each scenario to the observed solar-system $r$-process abundance distribution and the distribution of $r$-process elements in the early Galaxy.
The novel  conclusion of this paper is that one can possibly utilize the inherent shortcomings of the three  characteristic environments  to estimate the relative contributions of each  (weak, main, and fission recycling) to the final observed $r$-process abundances.

\section{Effect of Nuclear Closed Shells}
\begin{figure}[htbp]
	\begin{center}
  \includegraphics[scale=0.35,angle=-90]{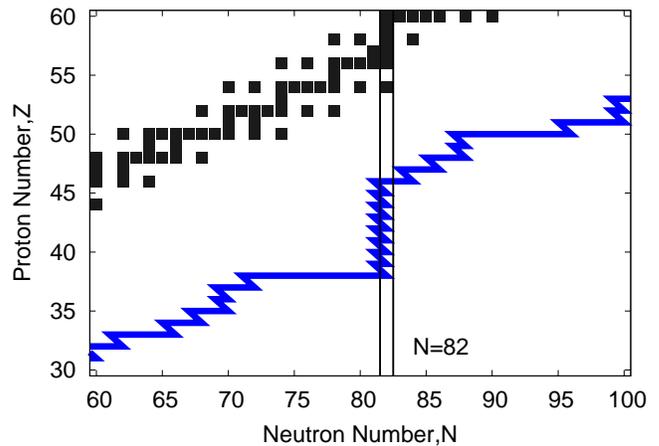}
    \caption{(Color online) Illustration of the (N,Z) path of $r$-process nucleosynthesis (blue line) for nuclei with $A \sim 90$ -- $150$ in the vicinity of the $N=82$ neutron closed shell  and Z=50 proton closed shell just before freezeout of the abundances in a typical main $r$-process (MHDJ) model.	 Black squares show the stable isotopes.}
\label{fig1}
	\end{center}
\end{figure}

Figure \ref{fig1} illustrates why the abundances below and above the $r$-process peaks are bypassed.  It shows an example of a typical calculated $r$-process path near the $N=82$ neutron closed shell just before freezeout when the  neutrons are rapidly exhausted and the abundances begin to beta decay back to the region of stable isotopes.   Neutron captures and photo-neutron emission proceed in equilibrium for nuclei with a neutron binding energy of about 1-2 MeV.  Above and below a closed neutron shell, however, this $r$-process path shifts abruptly toward the closed shell from below (or  away from the closed shell for higher nuclear masses).  This shifting of the $r$-process path toward the $N=82$ neutron closed shell causes isotopes with $N = 70-80$ ($A \sim$110-120) to be bypassed.  Similarly, the $A=$140-147 underproduction corresponds to the isotopes with proton closed shell $Z=50$ and $N\sim$90-97  ($A \sim$140-147).  These isotopes will also   be bypassed in the beta-decay flow as is evident on Figure \ref{fig1}.

We emphasize that this is not just an  artifact of the particular  mass model employed in this study.  Nearly all models in the current literature with a rapid freezeout  (including the MHDJ \citep{Nishimura06,Fujimoto07,Fujimoto08,Ono12,Winteler12,Nakamura14,Nishimura15}) show this underproduction if the final abundances are normalized to the abundance peaks.  Indeed, one is hard pressed to find any model for the main  $r$-process  (including  NDW models) in which this underproduction does not occur.   

One can of course contrive calculations to somewhat fill the dips on both sides of the second $r$-abundance peak.  Recently, for example, \cite{Lorusso15} were able to avoid the underproduction in a schematic high-entropy  outflow model by incorporating a summation of several  entropies.  In such a model one can fill in the dips similarly to the way we propose to do this by summing contributions from various  physical conditions.  As another example, calculations could fill the dips by using the ETFSI mass model as displayed in Fig. 7 in \cite{Nishimura06}).  However, these models do so at the cost of displacing the 2nd and 3rd peaks and/or  underproducing (or overproducing) abundances over a wide mass region between the second and third peaks.
This was also a consistent feature in the original realistic NDW models of \cite{Woosley94}.  Indeed this effect is apparent in almost every $r$-process calculation since the 1970s (cf.~review in \cite{Mathews85}).  

We note, however, that new attempts have been presented \citep{Kratz14} of  r-process calculations in a parameterized NDW scenario based upon the models of \cite{Freiburghaus99}.  Making use of new nuclear  masses and  beta-decay rates  from the finite-range droplet model
FRDM (2012) \citep{Moller12} it was shown  that the previous discrepancies near $A=120$ are  significantly diminished compared to the same calculation based upon the previous FRDM(1992) \citep{Moller95} nuclear properties. Hence, one must keep in mind that at least some of the apparent discrepancy may be due to the adopted nuclear input.

Although it has been speculated for some time \citep[e.g.][]{Woosley94,Pfeiffer01,Farouqi10} that this could be due to quenching of the strength of the shell closure or beta-decay rates near the closed neutron shell, this explanation is unlikely.    Recent $r$-process calculations \citep{Nishimura12} based upon new measurements \citep{Nishimura11} of beta-decay half lives near the $A=$130 $r$-process path have confirmed the absence  of shell quenching effects in the beta flow.  Moreover, recently  the first ever studies \citep{Watanabe13} of the  level structure of the waiting-point nucleus $^{128}$Pd  ($Z=$46, $N=$82 in Fig.~1) and $^{126}$Pd have been completed.  This study clearly indicates that the shell closure at the neutron number $N=$82 is fairly robust.  Hence, there is  absolutely no evidence of the hypothesized quenching effects in either the beta decay rates or nuclear masses.  One must suppose that some other resolution of this underproduction is necessary.  

One goal of this paper is, therefore,  to point out that a solution to the underproduction
of nuclei above and below the $r$-process abundance peaks can be obtained if one considers that a  fission recycling environment (e.g. NSMs) has contributed to the solar-system $r$-process abundance distribution in addition to the environments responsible for the weak and main $r$-process  (like CCSNe).  Indeed, this novel solution not only resolves this dilemma but can {\it quantify} the answer to the question of the relative contributions to the $r$-process abundances of the weak and main $r$-process environments (such as those due to CCSNe) vs. long-duration fission-recycling environments (such as NSMs) .

\section{Fission Recycling $r$-process}
For this paper we highlight the  possible role of  fission recycling  to account for the underproduction problem above and below the $r$-process peaks often found in  models for the main  $r$-process abundances.  For our purposes we employ a specific NSM model although we note that this is illustrative of any fission-recycling  $r$-process environment.  Nevertheless, the most natural current site for such fission recycling to occur is in the NSM models.    The ejected matter from  NSMs is very neutron-rich ($Y_e \sim 0.1$).  This means that   the $r$-process path proceeds along the neutron drip line all the way to the  onset of fission recycling.   As noted above, after a few cycles the abundances can become dominated by the fission fragment distributions and not as much by the beta-decay flow near the closed shells.  Hence, a very different mass distribution can ensue.  

In this regard we note that a number of recent studies \citep{Goriely11,Korobkin12,Goriely13,Wanajo14, Perego14} have indicated  that the $r$-process in NSMs can involve a distribution of neutron-rich environments.  Such models can produce a final abundance pattern that is similar to the solar-system $r$-abundances.  Here, even though we use the term NSM model, it is intended to refer to the portion of the ejecta that experiences fission recycling, while the other ejecta is similar to a NDW or MHDJ like model.  Hence, when we refer to the NSM model we really mean the ejecta that experiences fission recycling that  fills in the bypassed abundances produced in trajectories that produce the main $r$-process.   

An important point is that models including fission recycling effects   produce a final abundance pattern that is relatively insensitive to the  astrophysical uncertainties \citep{Korobkin12}, although the total (including non-recycling ejecta) abundances can be  sensitive to the detailed model.  

Nevertheless, the distribution of  nuclear fission products can affect the abundance pattern. 
Hence, one must carefully  extrapolate fission fragment distributions (FFDs) to the vicinity of the $r$-process path \citep[cf.][]{Martinez07,Erler12}.  We argue that by incorporating the expected broad distribution of fission fragments based upon phenomenological fits to observed FFDs, the effect of the neutron closed shells becomes smoothed out, thereby providing a means to fill in the isotopes bypassed in the main $r$-process.

For the present study we have made use of  self consistent  $\beta$-decay rates,  $\beta$-delayed neutron emission probabilities, and  $\beta$-delayed fission probabilities  taken from \cite{Chiba08}. The spontaneous fission rates and the $\alpha$-decay rates are taken from  \cite{Koura04}. In our $r$-process calculations,  $\beta$-delayed fission  is the dominant nuclear fission mode. Hence, for the most part other fission modes like neutron-induced fission can be neglected \citep{Chiba08}. 

To generate FFDs far from stability we have  made use of a semi-empirical  model \citep{Ohta07,Tatsuda08,Chiba08} that well reproduces the systematics of known fission fragment distributions.  This model can be naturally extrapolated to the required heavy neutron rich isotopes of the $r$-process.    As such it is a robust alternative means to predict yields from fission recycling.

A key ingredient of this model is that it can  account for   FFDs that can  either be single humped,  bimodal or even trimodal.  This is achieved by a weighted superposition of up to three Gaussian functions:
\begin{eqnarray}
f(A,A_{p})=\sum_{A_i} \frac{1}{\sqrt{2\pi}\sigma} W_i \exp\left(\frac{-(A-A_{i})^{2}}{2\sigma^{2}}\right) ~,
\label{eq:ffd}
\end{eqnarray}
where $A$ is the mass number of each  fission fragment, $A_{p}$ is that of the parent nucleus, $\sigma$ is the width of the three Gaussian functions, and the sum is over the possible fission fragment distributions, $i = L, H, M$, with 
\begin{equation}
A_{H}= \frac{(1+\alpha)}{2}(A_{p}-N_{loss})~~,
\end{equation}
\begin{equation}
A_{L}= \frac{(1-\alpha)}{2}(A_{p}-N_{loss})~~,
\end{equation}
and 
\begin{equation}
A_M =  \frac{(A_H + A_L)}{2}~. 
\end{equation} 
The factor $W_i$ is a weighting given by $(1 - \omega_s)$ for $i = L, H$ and $2\omega_s$ for $i = M$.  The quantities  $\omega_{s}$ and $\alpha$ are shape symmetry and mass-asymmetry parameters, respectively as defined below.  $N_{loss}$ is the number of prompt neutrons. 

For the present application we include the dispersion in the  FFDs ($\sigma=7.0$) and $N_{loss}=2$ from measured experiments on actinides. The adopted fission neutron emission  is an average value for all possible fission events.  We have run calculations in which this number $N_{loss}$ is varied from 2 to 8 and found that the results are nearly indistinguishable although a very small change is found below $A < 100$ and near the valley around $A = 180$.  
The atomic number and neutron number of each fission fragment is determined by the assumption that the  proton to neutron number  ratio is the same as that of the parent nucleus after correcting for prompt neutron emission, i.e.
$Z_{p}/N_{p}=Z/(N+N_{loss}/2)$. 

We have run calculations in which Gaussian width parameter $\sigma$ is varied from 4 to 14 and compared with the result with $\sigma=7$.  We  found that the rare-earth peak changes by only +20\%,-25\% so that the abundance decreases slightly for larger $\sigma$.  Although the abundances below $A < 100$ and near the valley around $A = 180$ increases as $\sigma$ increases, these changes do not change the overall distribution drastically, and the conclusions of this article are not affected by fixing $\sigma=7$.

The quantity $\alpha$ in Eqs.~(2)  and (3) is the average  mass-asymmetry parameter corresponding to the valley of the potential energy surface of the parent nucleus near the scission point for nuclear fission. 
This has been  calculated in the liquid drop model \citep{Myers99} with
shell energy corrections determined  \citep{Iwamoto76, Sato79} from the two-center shell model in the three-dimensional shape parameter space comprised of  $\alpha$, the distance between the centers of the two-harmonic oscillators $z$, and the deformation parameter of the fission fragments, $\delta$.  
The quantity $\omega_s$ is determined as $\omega_s = -0.2(V_s-V_a-2.0) $ for $V_s-V_a < 2.0$ MeV, and 
$\omega_s = 0$ otherwise, where  $V_s$ and $V_a$ denote the potential values at symmetric and asymmetric
valleys, respectively, at the fragment distance $z$ corresponding to scission.  This approximate
formula is derived to account for the  observed rapid change between  asymmetric- and symmetric-mass
distributions around $^{256}$Fm, i.e.
$V_s$ is adjusted relative to $V_a$ to reproduce the  observed mass distributions of the Fm isotopes with Eq. (1).

For illustration in the present  study  we have carried out $r$-process simulations in the fission recycling environment from the NSM outflow models of  \cite{Korobkin12,Piran13,Rosswog13}.  As an illustration of the main $r$-process we take abundances in the ejecta from the MHDJ model of  \cite{Nishimura12}.  For the  weak $r$-process we use yields from the NDW models of \cite{Wanajo13}.

Our adopted   NSM  outflow model is derived from   3D Newtonian smoothed-particle hydrodynamics (SPH)   \citep{Korobkin12,Piran13,Rosswog13}.  It gives qualitatively similar results to the fission recycling $r$-process yields calculated in 3D general relativistic SPH simulations \citep{Bauswein13,Goriely13} and the full 3D general relativistic simulations of  \cite{Wanajo14} in the  heavier mass region.  The details are different for lighter masses because of the broader range of neutron densities and electron fractions in the particle trajectories in those models.

We emphasize that the models run here can be considered as generic fission recycling models.  For specific  NSM models we  utilize the trajectories from the binary neutron star merger of two neutron stars with $M = 1.0$ M$_{\odot}$ each. Although 1.0 M$_{\odot}$ is not the typical neutron star mass \citep{Valentim11}, it has been shown  \citep{Korobkin12} that the resulting abundances are nearly independent of the  neutron star masses in the binary. 
The hydrodynamic simulations are based upon the SPH method of \cite{Rosswog09},  the equation of state (EoS) of \cite{Shen98a,Shen98b}, an opacity-dependent multi-flavor neutrino leakage scheme \citep{Rosswog03}, and Newtonian gravity. We use 30 available trajectories of neutron star merger ejecta to calculate the nucleosynthesis\footnote{Trajectories from http://compact-merger.astro su.se/}.  The ejected mass from this binary merger  is $\sim 0.01$ M$_{\odot}$ \citep{Korobkin12}.   After the end of the hydrodynamic simulation  at $t_{\rm fin}(\sim 15$ ms) the thermodynamic evolution can be  continued \citep{Rosswog14} as a free adiabatic expansion.

The reason for the use of these trajectories  is that they are publicly available and lead to robust fission recycling.    Moreover,  a main point of this paper is the importance of a fraction of $r$-process material that involves fission recycling.  If some fraction of the ejects in the NSM calculations as in \cite{Goriely11,Korobkin12,Goriely13,Wanajo14, Perego14} do not involve fission recycling, then this paper deals with the fraction of material in their models that leads to fission recycling, while the other trajectories would be absorbed into what we label as other components.   

In contrast to CCSNe, baryons participating in the $r$-process constitute a large fraction of the total mass-energy in the NSM ejecta. Thus, the nuclear energy released by the $r$-process nuclear reactions must be  included after $t_{\rm fin}$ by an entropy source term,
\begin{equation}
dS=-\epsilon_{\rm th}\sum _{i}\left( {m_{i}c^{2}}/{k_{B}T}+{\mu_{i}}/{k_{B}T}\right)dY_{i}~,
\end{equation}
where a heating efficiency parameter $\epsilon_{\rm th} \approx 1$ is introduced  \cite{Korobkin12} to account for neutrino energy losses.  
 
The nucleosynthesis calculations were started at a temperature  $T=9.0 \times 10^9$ K.  At this point all nuclei are in nuclear statistical equilibrium, and the composition is completely determined from the density and charge-per-baryon $Y_e$ of the material ejected from the neutron stars.  At this point the material almost  entirely consists  of free neutrons plus some heavy seed nuclei with A$\approx$70.  

As the temperature and density decrease, however, the material is evolved using an updated version of the  nuclear network code of  \cite{Terasawa01}.  The neutron radiative capture rates are as summarized in \cite{Terasawa01}, however, for both the weak and main  hot $r$-process considered here, the  abundance patterns mainly depend on the nuclear masses and beta-decay rates but not on the radiative neutron-capture rates. This is because the system proceeds in $(n,\gamma)$ equilibrium until a rapid freezeout of the neutron abundance.  On the other hand, the  fission-recycling NSM $r$-process considered here depends on the radiative neutron-capture rates because, when the temperature is low, the $(n,\gamma)$ and  beta-decay rates (in the so-called "cold $r$-process") determine the final abundances.  In the fission recycling  model adopted here, the $r$-process path terminates in a region where beta-induced  fission is much faster than the neutron-induced fission so that  the $r$-process is always terminated by beta-induced fission.  We note, however, that this depends upon the treatment of fission barriers and a different treatment \citep[e.g.][]{Korobkin12} can result in a different mode of fission  termination. 

 Once the $r$-process path fissions, we utilize  the fission fragment distributions given in \cite{Ohta07} and also the nuclear masses from the KTUY model \citep{Koura05}.   The fission barriers are extracted from the same KTUY model.  However, since the KTUY model treats only the symmetric fission, we adopted here the two-center shell model to allow more general fission fragment distributions.  
The KTUY model has been shown within the GT2 theory to reproduce recent measurements of beta-decay half-lives of exotic neutron-rich isotopes \citep{Nishimura11}. In a separate forthcoming paper we will summarize a detailed comparison of the predictions of this model with known FFDs.

We also note that there have been numerous studies \citep[e.g.][]{Otsuki03,Pfeiffer01}  of the sensitivity of this type of paradigm on various nuclear physics parameters.  However, almost  all MHDJ (or NDW) models (without shell quenching) show the abundance deficiencies on either side of the closed $r$-process abundance peaks.   Moreover, the NDW and MHDJ supernova models often involve little or no fission recycling.  As such, they do not depend on the details of fission rates and fragment distributions. 

We note, however,  that our NSM calculation (as shown below in Fig. \ref{fig2}) produces a different  abundance pattern than that of previous  NSM studies \citep{Goriely11,Korobkin12,Goriely13}, especially  in the region spanning between  the  2nd and 3rd $r$-process peaks. There are two reasons for this difference: 1) The fission fragment distributions (FFDs); and 2) the number of fissioning nuclei contributing to fission recycling and the freezeout of the $r$-process distribution.  

Regarding the FFDs,  it  has often been noted  \citep[e.g.][]{Goriely11,Korobkin12,Goriely13} that the elemental abundances from NSM calculations depend strongly on the  FFD model.  Admittedly this is a major uncertainty in all calculations of fission recycling in the $r$-process.  As noted above, our FFD model is based upon the KTUY model plus a two-center shell model to predict both symmetric and asymmetric FFDs with up to three components.  As such,  fissile nuclei in our approach can span a wide mass range (A=100-180) of fission fragments.  This is illustrated in the upper panel of Fig. \ref{fig2} that shows the final abundance distribution compared with the FFDs  of three illustrative nuclei.    

On the other hand, the models of \cite{Korobkin12} are mostly based upon a simple two fragment distribution as in \cite{Panov01} (or alternatively the prescription of \cite{Kodama75}).  The assumption of only two fission daughter nuclei  tends to place a large yield near  the second $r$-process peak leading to a distribution that looks rather more like the solar $r$-process abundances.  In contrast, the FFDs of \cite{Goriely13} are based upon a rather sophisticated SPY revision \citep{SPY} of the  Wilkinson fission model \citep{Wilkins76}.  The main ingredient of this model is that the individual potential of each fission fragment is obtained as a function of its axial deformation from tabulated values. Then a Fermi gas state density is used  to determine the main fission distribution.   This leads to a multiple hump FFD similar to that considered here, but even with  up to four humps.  Although this arguably represents a more fundamental approach than that employed in the present work, we prefer the phenomenological FFD approach here as an alternative means to estimate fission yields far from stability.  

An even more important  difference between  the present work and that of previous studies is the termination of the $r$-process path and the number of fissioning nuclei that contribute to   fission recycling and the  freezeout of the $r$-process abundances.  
The $r$-process path in our NSM calculations proceeds rather below the fissile region until nuclei with $A \sim 320$, whereas the $r$-process path in \citep{Goriely13} terminates at $A \approx 278$ (or for a maximum $\langle Z \rangle$ for \citep{Korobkin12}). Moreover, we find that only $\sim10$\% of the final yield comes from the termination of the $r$-process path at N = 212 and Z = 111,  while almost 90\% of the $A=160$ bump shown in Fig. \ref{fig2} comes from the fission of  more than 200  different parent nuclei mostly via beta-delayed fission. 
This is in contrast to the yields of \cite{Goriely13} that are almost entirely due to a few  $A \approx 278$ fissioning nuclei with a characteristic four hump FFD.  This is the reason why they obtain a solar-like  $r$-process like distribution. 

To illustrate this point, in the lower panel of Fig. \ref{fig2} we compare the yields of our model  with a calculation in which we assume that the $r$-process path is terminated by symmetric fission of  nuclei with $A=285$.  Clearly, in this case a solar-like distribution is obtained similar to that of Refs. \citep{Goriely11,Korobkin12,Goriely13}.  This highlights the importance of detailed fission probabilities   along the $r$-process path.

Finally, we note that the apparent suppression of the  the 3rd $r$-process peak in our final abundances relative to that of other works is caused by the large increase in  the rare earth elements resulting from the FFDs of repeated fission recycling.

\begin{figure}
 \centering
 \includegraphics[scale=0.65]{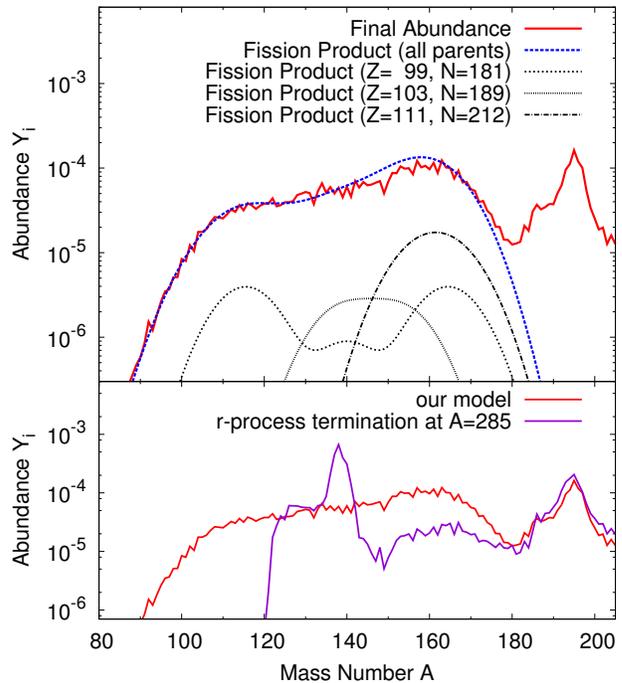}
  \caption{(Color online) Illustration of the impact of fission yields and fission recycling  on the final $r$-process abundances. Upper panel shows the relative contributions for 3 representative nuclei compared with the final abundance distribution.    The lower panel shows the same final $r$-process yields  compared with the distribution that would result if fission recycling were only to occur from parent nuclei at the termination of the $r$-process path at $A=285$.    }
 \label{fig2}
\end{figure}

\section{relative $r$-process  contributions}
\begin{figure}
 \centering
 \includegraphics[scale=0.35,angle=-90]{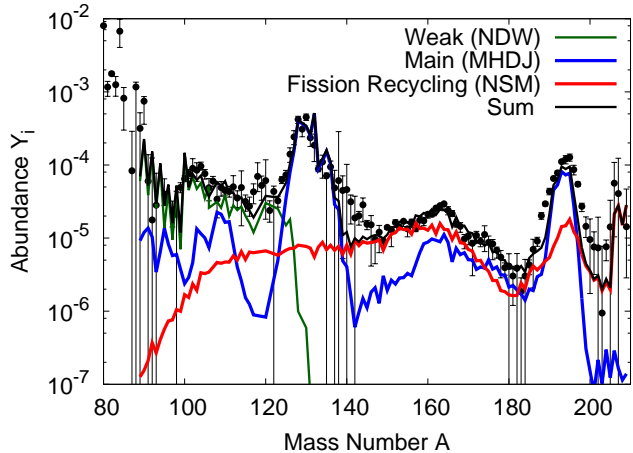}
  \caption{(Color online) Average final abundance patterns for the fission recycling environment of NSM  (red line), the main $r$-process abundances from the MHDJ model (blue line) and  weak $r$-process (green line) from the NDW.  These are compared with the  observed \citep{Goriely99} $r$-process abundances in the solar system (black dots).  The thin black line shows the sum of all contributions.}
 \label{fig3}
\end{figure}

Figure \ref{fig3} shows the main result of this paper.  The red line on Figure \ref{fig3} shows the result of our fission recycling nucleosynthesis simulation summed over all trajectories of material ejected from the binary NSM model adopted here.  This is compared with the abundances in the ejecta from the main   $r$-process (blue line) from  the MHDJ model of \cite{Nishimura12}, and also  the NDW weak $r$-process abundances (green line) produced in the NDW  from the 1.8 M$_\odot$ supernova core model of \cite{Wanajo13}.  

The key point of this figure is the important role that each process plays in producing the total abundance pattern of solar-system $r$-process abundances [black dots with error bars \citep{Goriely99}].
The total abundance curve from all processes is shown as the black line on Figure \ref{fig3}.  The  weighting factor $f_{Fission}$ was determined from a normalization to isotopes near A=145-155 for the fission recycling (NSM) model. The factor  $f_{Weak}$  was determined from  a fit to light isotopes near A=100  for the NDW model.   The MHDJ yields were  normalized to the second $r$-process peak.
The best fit (black) line in Figure \ref{fig3} is for $f_{Fission} = 0.16$ and $f_{Weak} = 4.3$, or roughly 79\% weak, vs.~18\% main, and $\sim 3$\% fission-recycling contributions with  some uncertainty  in the different models as noted above.  Nevertheless, these estimated relative contributions  are at least consistent with roughly estimated Galactic yields as described below.

Of particular relevance to the present study is that the underproduction of nuclides above and below the $A=130$ $r$-process peak shown by the blue line is nearly accounted for by the fission recycling (NSM) and weak $r$-process (NDW) models.
The final NSM $r$-process isotopic  abundances from our adopted model for fission yields exhibit a very flat pattern due to several episodes of fission cycling. 
Thus, we find that fission recycling has the potential to resolve most of the underproduction problems for the elements just below and above the abundance peaks in models of the main $r$-process.  
The remaining underproduction below the $A=130$ peak is most likely due to the weak $r$-process as illustrated on Figure \ref{fig3}.

The main point of this paper is that one can deduce the relative contributions of each $r$-process model based upon their relative shortcomings.  However, it is important to ask whether the inferred fractions, of $\sim$79\% NDW, $\sim$18\% MHDJ, and $\sim$3\% NSM  are plausible.   

Although there are many uncertainties in the astrophysical and galactic chemical evolution parameters \citep{Argast00, Komiya14}, it is worthwhile to estimate  weight parameters $f_{Fission}$ and $f_{Weak}$  from observed Galactic event rates and expected yields.  In particular we write
\begin{equation}
f_{Fission} \approx \frac{\rm R_{NSM} {\rm M_{r, NSM}} }{ \epsilon_{MHDJ} {\rm R_{CCSN}} {\rm M_{r, MHDJ} }}~~,
\end{equation} 
 and 
\begin{equation}
f_{Weak} \approx  \frac{\rm R_{CCSN} {\rm M_{r, Weak} }}{\epsilon_{MHDJ} {\rm R_{CCSN}} {\rm M_{r, MHDJ} }}~~,
\end{equation}
where $\rm M_{r, NSM}$, $\rm M_{r, MHDJ}$, and  $\rm M_{r, Weak}$ are the ejected masses of $r$-elements from the NSM, MHDJ, and NDW  $r$-process models, respectively, while $\rm R_{CCSN}$ and $\rm R_{NSM}$ are the corresponding  relative Galactic event rates of CCSNe and NSMs.  

The ejected mass of $r$-process elements in the models of \cite{Wanajo13} is $\approx 2 \times 10^{-5} $ M$_\odot$ and nearly independent of assumed core mass.  The quantity $\epsilon_{MHDJ}$ is the fraction of CCSNe that result in magneto-rotationally driven jets. This  was estimated  in \cite{Winteler12} to be $\sim 1$\% of the core-collapse supernova rate based upon the models of \cite{Woosley06}.  However this is probably uncertain by at least a  factor of two.  Indeed, the fraction could be larger as most massive stars are fast rotators and the conservation of magnetic flux should often  lead to high magnetic fields in the newly formed proto-neutron star.  Hence, this fraction could easily range from  $\sim 1$ to $\sim 5$\% which incorporates the $\sim 1$\% fraction of observed magnetars compared to normal neutron stars.  [We treat this as a lower limit because some fraction of observed normal neutron stars may have had a larger magnetic field in the past.]
The mass of synthesized $r$-process elements from  MHDJs is estimated to be 6$\times$10$^{-3}M_{\odot}$ \citep{Winteler12} while that of a typical  binary NSM is expected to be $2 \pm 1\times10^{-2}M_{\odot}$ \citep{Korobkin12}.
If the Galactic neutron star merger rate is $80^{+200}_{-70}$ Myr$^{-1}$ \citep{Kalogera04}, and the Galactic supernova rate is, $1.9 \pm 1.1 \times 10^4$ Myr$^{-1}$ \citep{al26},  then one should expect $f_{Fission} \sim 0.6 \pm 0.4 $  and $f_{Weak} \approx  8 \pm 6$ corresponding to  relative contributions of  $\sim 80$\% weak, $\sim 10$\% main and $\sim 10$ \% fission recycling.  Thus, although there are large uncertainties, these fractions are  plausibly consistent with our fit parameters.  This suggests that such a fit  may be a way of constraining the relative contribution of NSMs and CCSNe to solar-system material.
 
We note, however, that other NSM
calculations predict about $10^{-4}$ to  $10^{-2}$ M$_\odot$ of $r$-process material to be ejected
\citep[e.g.][]{Hotokezaka13, Bauswein13}. Adopting a value of
$10^{-3}$ M$_\odot$ could lead to $f_{Fission} \sim 0.02$, i.e. about an order of magnitude below that suggested in our fit to Figure \ref{fig3}.  

Of course, this needs to be better quantified  in more detailed chemical evolution \citep{Cescutti14,Cescutti15,Tsujimoto14a,Tsujimoto14b,Komiya14,Ishimaru15,Wehmeyer15} and chemodynamical studies \citep{Shen15,vandeVoort15} along with better $r$-process hydrodynamic models \citep{Winteler12,Perego14,Rosswog14,Wanajo14,Goriely15,Just15,Nishimura15}.  Nevertheless, based upon the models adopted here, the inferred division of $r$-process contributions remains at least plausible.

\section{universality of $r$-process elemental abundances}
In the above we have not discussed a very important clue to the origin of $r$-process abundances.  It is by now well established \citep{Sneden08} that the elemental abundances in many metal-poor stars show a pattern that is very similar to that of the solar-system $r$-process distribution, particularly in the range of $55 < Z< 70$.   This however, can pose a difficulty  \citep{Mathews92,Argast00} for NSM models (either in the present work or in other studies).   That is because  metal-poor stars are thought to have arrived very early in the history of the Galaxy, whereas NSMs require a relatively long gravitational radiation orbit decay time prior to merger ($\sim 0.1$ Gyr).  
Whatever the situation, it is of value to examine the impact of the possible late arrival of fission recycling material  on the $r$-process elemental abundance distribution in metal-poor stars.  

\begin{figure}
 \centering
 \includegraphics[scale=0.35,angle=-90]{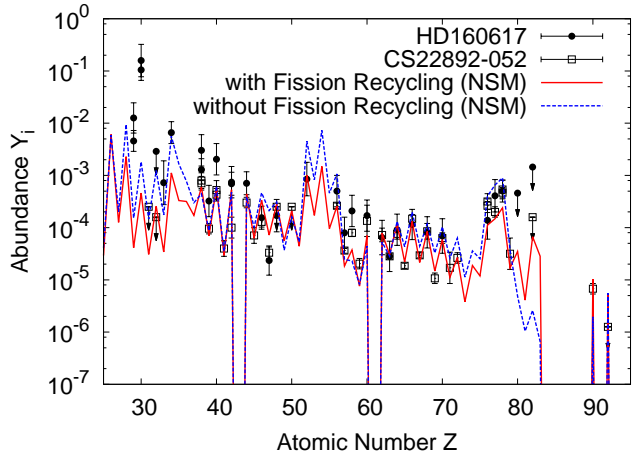}
  \caption{(Color online) Average final  elemental  abundances for the total sum from Fig.~\ref{fig2} (solid line) and the contribution without NSMs (dashed line).  These are compared with the  observed elemental  $r$-process abundances in two well-studied metal-poor $r$-process enhanced stars, HD1601617 (filled circles; \cite{Roederer12}) and CS22892-052 (open squares; \cite{Sneden03}).  The curves are arbitrarily normalized at europium (Z=63).}
 \label{fig4}
\end{figure}

Figure \ref{fig4} shows the elemental abundance distribution calculated in two scenarios, i.e.~with and without the fission recycling yields of NSMs.   These are compared with the  observed elemental  $r$-process abundances in two well-studied metal-poor $r$-process enhanced stars, HD1601617  \citep{Roederer12} and CS22892-052 \citep{Sneden03}. Here, we note that there is little distinction between the two curves (although the fit is slightly better when the fission recycling yields  are included).  The reason for this insensitivity is that the fission recycling environment  only contributes about 3\% to the total $r$-process abundance.  Although this yield is  important to fill in the  isotopic abundances above and below the $r$-process peaks, and also to make the rare-earth bump near A=160, there is little apparent difference in the elemental abundances with or without neutron star mergers.  Among other things, this is because the region below the peak ($Z\sim 50$) is poorly sampled, and moreover, summing over isotopes to produce elemental averages somewhat washes out the underproduction above and below the $r$-process mass peaks.  Hence, the elemental $r$-process abundances in metal poor stars do not clearly require that  fission recycling  occurred early in the Galaxy.  

We do note, however that the dispersion in the stars themselves for the lightest elements ($30 < Z < 50$) is suggestive that not all CCSNe contribute both a weak  and main  $r$-process.  This is consistent with the expectation that the NDW could occur in all CCSNe while  the main $r$-process from the MHDJ will only occur in a limited fraction of CCSNe, i.e.~those with rotation and strong magnetic fields.

\section{Discussion}

The fits to the abundance distribution (e.g. Figure \ref{fig3})  are as good as or better than  most  models in the literature.  Nevertheless,  it is worthwhile, to address some of the detailed deficiencies in both Figures \ref{fig2}  and \ref{fig3}.  For example, although the $r$-process peaks at A=130 and 195 along with the rare-earth peak region A = 145-180 in Figure \ref{fig3} are remarkably well reproduced,  there are some differences just above the main $r$-process peaks in the regions of A=140-145 and 200-205.   We note, however, that these isotopes have the largest uncertainties in the $r$-process abundances themselves as is visible on Figure \ref{fig3}.  Hence,  these discrepancies may simply reflect the abundance uncertainties, although the possibility remains of a  shortcoming in the models for these isotopes.  

Similarly, in Figure \ref{fig4} there is an  underproduction of elements at  Z=58 and 60. The  abundance of Ce (Z=58) in Figure 4 is well determined observationally for CS22892-052 as follows: $\log{\epsilon}({\rm Ce})=  -0.50\pm 0.07$ \citep{Sneden03} and  $=-0.38\pm 0.08$ \citep{Honda04}. This  corresponds to the deficient isotopes   with A=140 and  142  in Figure \ref{fig3}. However, the odd elements with Z= 57 (La) and Z=59 (Pr) are reproduced.  This suggests that the odd-even effect in the region of lanthanide elements may be  underestimated in the mass model employed here.  Nevertheless, the main point of this paper is not to give a precise reproduction of $r$-process elemental abundances but rather to demonstrate the possibility that fission recycling supplements the underproduced elements.  Clearly, a better understanding of the nuclear uncertainties within this context is still needed. 

 We also note that there is a possible deficiency of Pb (Z=82) in Figure \ref{fig4}.  This, however, may relate to observational uncertainties.  There are two measured Pb abundances for  CS22892-052 in \cite{Sneden03}.
One was  a ground-based measurement,  while the other was obtained with  {\it HST}.  However, both of these values should be considered upper limits.
In \cite{Sneden03} it was noted that the suggested detections of the two Pb I lines in the ground-based spectra should be nearly 10 times weaker than the $\lambda= 2833.05$ line, that could not be  detected in the HST spectrum. Hence, the derived Pb abundance upper limit from the $\lambda =2833$ line is probably  more reliable than the abundances determined from the questionable detections of the other two Pb I lines. 
Thus, one should abandon the  Pb abundance of $\log{\epsilon({\rm Pb})}=0.05$ from the ground-based observation in favor of  $\log{\epsilon({\rm Pb})}<-0.2$ from the HST observation.  We also  note that the more recent observation of \cite{Roederer09} also obtains $\log{\epsilon({\rm Pb})}<-0.15$.  These upper limits are  consistent with our calculation.

Another  issue worthy of discussion is that of Th (Z=90) and U (Z=92) production in Figure \ref{fig4}.  Th has been observed in a number
of metal-poor stars and U in a few.  This indicates that the $r$-process mechanism at work in the early Galaxy could produce the actinide elements and beyond.
Although one tends to think that the production of actinide elements requires a fission-recycling  $r$-process, in fact the production of Th and U is possible even in models that do not lead to fission recycling.    For example, the  MHDJ models with strong magnetic fields in \cite{Nishimura15}  could produce Th and U in as much as their solar abundances.  On the other hand, 
MHDJ models with weak magnetic fields tend to produce actinides below solar abundances.  Hence, the observation  of Th and U in metal-poor stars constrains  the early astrophysical environment,  but does not necessarily require that a fission-recycling  $r$-process (such as the NSM model) contributed to metal-poor stars.   

\section{Conclusions}
In summary, we have considered the relative contributions of three generic  $r$-process environments to the solar-system $r$-process abundances and the abundances in $r$-process enhanced metal-poor stars.   These environments are discussed in the context of   neutron star mergers, neutrino driven winds and magnetohydrodynamically driven jets, although these environments  should be considered as illustrative and  not definitive of the specific $r$-process environments.  Nevertheless, based upon our adopted fission fragment distributions we find that
the relative contributions from each environment has the possibility of explaining a unique feature of the $r$-process abundances.  Moreover,   the deduced relative contributions are plausibly consistent with galactic chemical evolution considerations.  

Clearly, more work along this line is required to explain details.  Nevertheless, we suggest that the possibility that all three general environments occur in detectable amounts in the $r$-process distribution should be taken seriously in future investigations of the origin of $r$-process nuclides.  

\vskip .1 in
We thank M. Famiano, K. Nakamura, and J. Hidaka for useful discussions, and also N. Nishimura for his nucleosynthesis data of MHDJ from Ref. Nishimura et al. (2012), and S. Rosswog  for providing NSM trajectories.  We also thank  T. Tachibana and H. Koura for providing nuclear reaction data. Numerical simulations were carried out at the Center for Computational Astrophysics, National Astronomical Observatory of Japan.  
Work at NAOJ was supported in part by Grants-in-Aid for Scientific Research of JSPS (26105517, 24340060). 
Work at the University of Notre Dame is supported 
by the U.S. Department of Energy under
Nuclear Theory Grant DE-FG02-95-ER40934.

\end{document}